# Probing atomic-scale symmetry breaking by rotationally invariant machine learning of multidimensional electron scattering

Mark P. Oxley,[1] Maxim Ziatdinov,[1,2,1] Ondrej Dyck,[1] Andrew R. Lupini,[1] Rama Vasudevan,[1] and Sergei V. Kalinin[1,2]

[1] Center for Nanophase Materials Sciences, Oak Ridge National Laboratory, Oak Ridge, TN 37831, USA

[2] Computational Sciences and Engineering Division, Oak Ridge National Laboratory, Oak Ridge, TN 37831, USA

The 4D scanning transmission electron microscopy (STEM) method has enabled mapping of the structure and functionality of solids on the atomic scale, yielding information-rich data sets containing information on the interatomic electric and magnetic fields, structural and electronic order parameters, and other symmetry breaking distortions. A critical bottleneck on the pathway toward harnessing 4D-STEM for materials exploration is the dearth of analytical tools that can reduce complex 4D-STEM data sets to physically relevant descriptors. Classical machine learning (ML) methods such as principal component analysis and other linear unmixing techniques are limited by the presence of multiple point-group symmetric variants, where diffractograms from each rotationally equivalent position will form its own component. This limitation even holds for more complex ML methods, such as convolutional neural networks. Here, we propose and implement an approach for the systematic exploration of symmetry breaking phenomena from 4D-STEM data sets using rotationally invariant variational autoencoders (rrVAE), which is designed to disentangle the general rotation of the object from other latent representations. The implementation of purely rotational rrVAE is discussed as are applications to simulated data for graphene and zincblende structures that illustrate the effect of site symmetry breaking. Finally, the rrVAE analysis of 4D-STEM data of vacancies in graphene is illustrated and compared to the classical center-of-mass (COM) analysis. This approach is universal for probing of symmetry breaking phenomena in complex systems and can be implemented for a broad range of diffraction methods exploring the 2D diffraction space of the system, including X-ray ptychography, electron backscatter diffraction (EBSD), and more complex methods.

[1] ziatdinovm@ornl.gov
[2] sergei2@ornl.gov

Functionalities of materials including ferroics,[1,2] superconductors,[3] and charge density wave systems[4] are governed by the physics of symmetry breaking phenomena. In systems with long-range discrete translation symmetries, these behaviors are readily amenable to neutron and X-ray scattering, providing insight into the minute details of atomic structure, electronic density distribution, and elastic and inelastic vibrational properties.[5,6] In these systems, the long-range periodicity allows integrating the behaviors over multiple unit cells. Similar approaches can be extended to ordered 2D systems such as surfaces and interfaces, as accessed via low-energy electron diffraction or surface X-ray methods.[7,8]

However, this approach offers only limited applicability to materials such as nanoscale phase-separated oxides, ferroelectric relaxors and morphotropic phase boundary systems, incommensurate charge- and spin density wave systems and, more generally, systems with non-uniform ground states. Similarly, the local mechanisms describing the interplay between chemical disorder, including both lattice-preserving substitution and lattice breaking structural defects, and physical functionalities are often unknown. In all these cases, the lack of long-range translational symmetry limits the applicability of classical scattering techniques and requires development of methods for probing correlated disorder.

At the same time, the last several years have seen an exponential growth of atomic-scale electron diffraction in scanning transmission electron microscopy (4D-STEM). The fast electrons in the electron probe are deflected by the electric field within the crystal. Negatively charged electrons are attracted to positively charged nuclei, which are screened by the surrounding electrons, meaning they contain sub-atomic scale components. This variation is most clearly seen in diffraction space, where the center-of-mass (COM) of the convergent beam electron diffraction (CBED) pattern is deflected toward the nuclei. Practically, the atomically sized focused electron beam is used to collect the local (2D) diffraction patterns over a dense spatial grid of (2D) points, producing the 4D-STEM data sets. A unique aspect of this method is that the size of the probe can be below the distance between the scatterers, resulting in very complex local diffraction patterns and encoding minute details of the local scattering potential.

Originally, 4D-STEM in its modern form was proposed by Rodenburg as an approach to achieve high spatial resolution,[9,10] enabling a practical embodiment of the ptychographic idea of Hoppe.[11,12] However, there were two main difficulties that prevented the widespread adoption of these methods. First, a practical problem was that CCD cameras were not fast or sensitive enough to keep up with the speed of the STEM probe, resulting in long acquisition times creating sample damage and stability problems. The second main problem was that the data sets were too large for existing computer infrastructure and the amount of computation required made it prohibitively expensive. Both of these difficulties have been addressed over last 4-5 years. Modern computers and their associated storage and data-handling capabilities have improved dramatically in accordance with the well-known Moore's law. Electron detection capabilities have grown both evolutionarily with incremental improvements in conventional designs and revolutionarily with the advent of direct-electron detectors.[14-17]

Methods other than ptychography have been developed to analyze scanning nanodiffraction data. The position averaged CBED (PACBED) approach has been used primarily to determine specimen thickness.[18] PACBED has recently been enhanced by the application of deep convolution neural networks to automatically analyze the data sets. Differential phase contrast (DPC) in the STEM was originally proposed in the early 1970s[20] and was recently implemented using segmented detectors.[21] The development of high-speed electron detectors has allowed DPC-STEM to be readily applied. By determining the deflection of the COM of the CBED pattern as a function of probe position, insight can be gained about the local charge densities and fields[22] or alternatively the electron scattering potential.[23]

Despite these initial advances and the well-recognized promise of 4D-STEM for the sub-atomic scale exploration of materials properties, progress has been stymied by a lack of analysis tools to convert the 4D-STEM data sets into physically relevant parameters. The vast majority of the work presently relies on using a simple COM. Alternatively, a number of approaches using linear unsupervised dimensionality reduction methods such as principal component analysis (PCA) and non-negative matrix factorization (NNMF) and clustering techniques have been explored and recently have become part of open-source platforms.

The applicability of linear separation methods for analysis of 4D-STEM data sets is limited, stemming from the intrinsic symmetries of the atomic lattice as illustrated in Fig. 2. Linear unmixing methods such as PCA will separate Ronchigrams that differ by in-plane rotation only, creating multiple components describing rotational states of nominally identical objects. Similarly, conventional deep neural network architectures employing rigid convolutional layers combined with the distortions and deformations that are universally present in the imaging system and the mesoscale strain fields in the material will give rise to a very large number of weakly meaningful components that do not allow for direct physical interpretation. Here, we propose an approach for the analysis of 4D-STEM data based on rotationally invariant autoencoders. In general, variational autoencoders (VAEs) are one of the primary classes of generative ML models that seek optimum representation of input high-dimensional data sets in terms of a small number of latent variables. More specifically, VAEs belong to a family of directed latent variable probabilistic models that can infer hidden structure in the underlying data.[24,25] We assume that each observed data point, $x_i$, is generated in a non-linear way by some latent variable, $z_i$, and that the joint probability density of the generative model can be expressed as:

$$p(\mathbf{x}, \mathbf{z}) = \prod_{i=1}^{N} p_\theta(\mathbf{x}_i|\mathbf{z}_i) p(\mathbf{z}_i),$$

where $\theta$ is a global parameter that all datapoints depend on. In VAE, one introduces a variational family of distributions that approximate the true, but intractable posterior distribution, $q_\phi(\mathbf{z}|\mathbf{x}) \approx p_\theta(\mathbf{z}|\mathbf{x})$. The latent variable model is then learned by maximizing the evidence lower bound (ELBO) with respect to the model parameters, $\theta$, and the variational parameters, $\phi$, for any given datapoint $\mathbf{x}$. In practice, $q_\phi(\mathbf{z}|\mathbf{x})$ and $p_\theta(\mathbf{x}|\mathbf{z})$ are parameterized by deep learning networks, usually referred to as the encoder and decoder, where $\phi$ and $\theta$ are trainable weights optimized by stochastic gradient descent (SGD) algorithms. Unlike linear methods such as PCA, VAEs often allow for much more efficient representation of rotationally equivalent forms.

Here, we combine the intrinsic parsimony of VAE with rotational symmetry, allowing for efficient encoding of equivalent units at different rotations. To account for rotational invariance, we adapted the approach of Bepler *et al.*[26] who showed that one can disentangle latent variables associated with image content and those associated with image rotation by parameterizing the decoder as a function of the spatial coordinates of the image. In this case, a single forward pass consists of (i) the encoder outputting parameters of a probabilistic distribution (chosen to be a diagonal Gaussian), (ii) the decoder generating a latent vector by sampling from the encoded distributions, followed by (iii) splitting the latent vector into the part associated with image content and the part associated with image coordinates and performing a 2D rotation of the latter by a random angle, and finally (iv) passing both the transformed coordinates and the sampled image latent vector through the decoder neural network to reconstruct the original output. This process is illustrated graphically in Fig.1. The encoder and decoder weights are optimized jointly with the ELBO loss function consisting of two Kullback-Leibler divergence terms,[27] one for image content and the other for rotations in addition to a reconstruction loss term using the Adam extension[28] of SGD with a learning rate of 0.0001. Both encoder and decoder have a simple multi-layer perceptron structure with two layers and 128 neurons per each layer activated by a *tanh()* function. The nature of the VAEs dictate that both feature and target data are the encoded data sets.

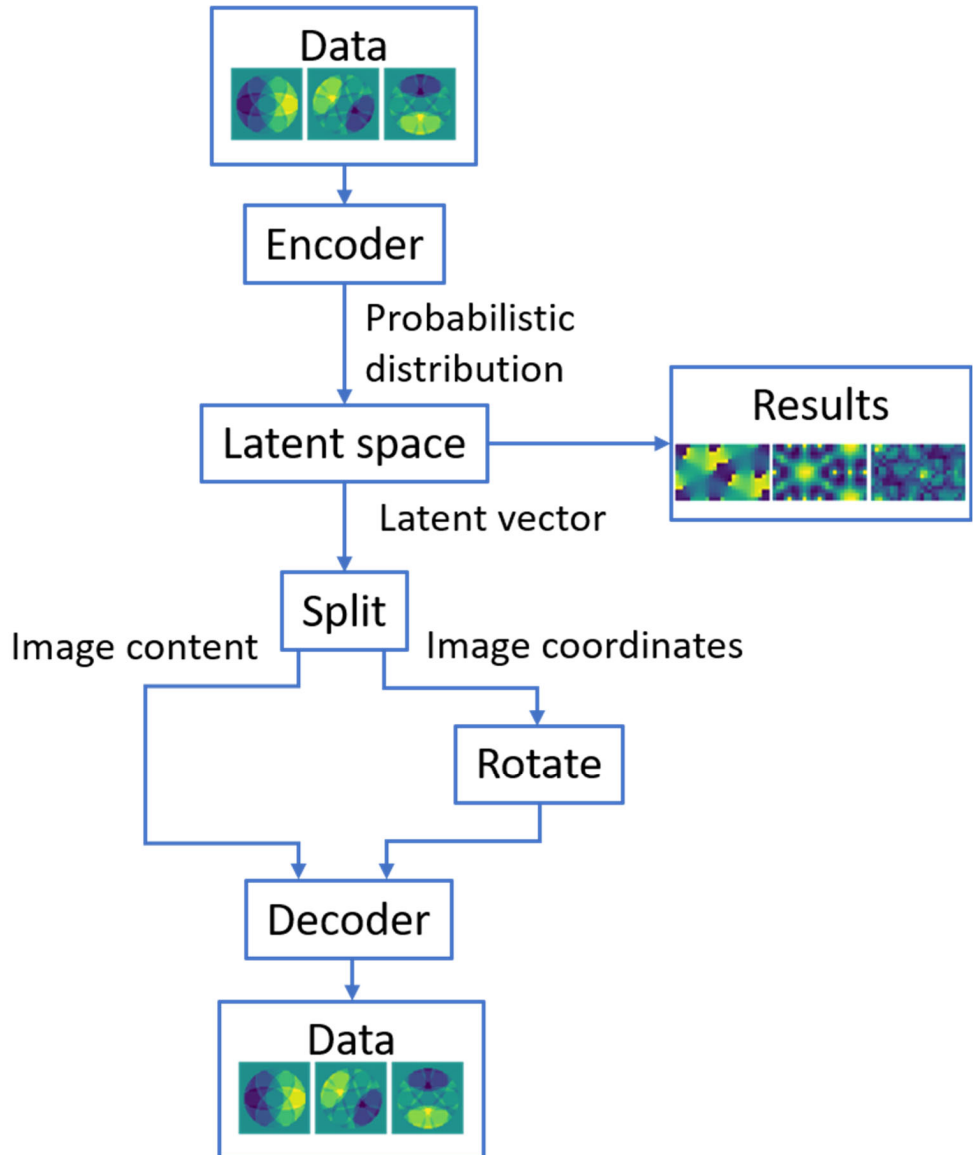

**Figure 1.** Graphical representation of the rrVAE algorithm as described in the text.

Figure 2 shows a plot of the simulated CBED patterns as a function of probe position for 60 kV electrons incident on graphene. An aberration-free probe with a 31 mrad probe forming aperture is used, which is chosen to be close to that used in the experiment [29]. The CBED patterns are normalized by subtracting the mean CBED intensity over all positions. This process is also helpful in the subsequent rotationally invariant VAE (rrVAE) analysis (which is like subtracting the mean in the PCA). The degree of deflection depends on the closeness of the probe to the atomic site and the electric fields of the other atoms, which leads to many CBED patterns with similar shapes but different rotations. It is this variation that is used in the COM methods to reconstruct the electric fields and related quantities. Hence, the relevant question is whether rrVAE allows us to determine the same physical properties and perhaps provide additional insights in the structure of the 4D-STEM data sets.

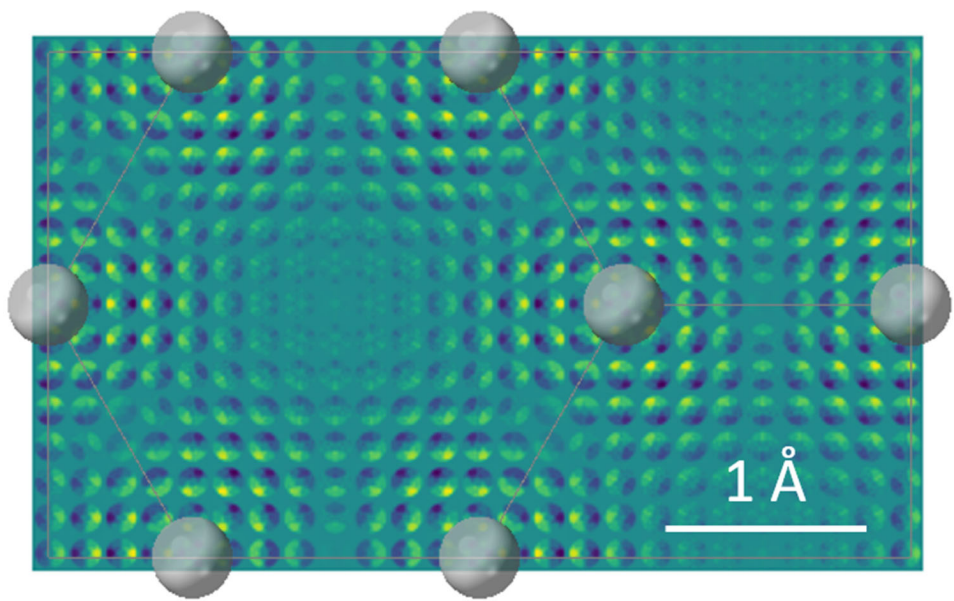

**Figure 2.** Simulated CBED patterns as a function of probe position for 60 kV incident electrons on graphene overlayed on the atomic positions. An aberration-free probe with a probe forming aperture of 31 mrad was assumed with resulting CBED patterns having a diameter of 62 mrad. No incoherence was added at this stage.

For rrVAE training, it is important to have a consistent stopping criterion similar to most iterative processes. For the specific configuration used, the convergence of the rrVAE is examined in Supplementary Fig. S1, using the simulated dataset above. The training loss decreases rapidly at first and then gradually flattens and reduces slowly in a monatomic fashion. While it might (naively) seem that more iterations would provide a better result, the results actually degrade if too many iterations are performed. In many cases, the latent spaces appear closely related to the COM deflection map shown in Figure 3 (d). In order to provide a robust measure of the correlation between the latent spaces and the COM deflections, we use the Pearson correlation coefficient or the Pearson r factor that ranges in value between 1 and -1, with 1 being a perfect positive linear correlation and -1 being a perfect negative linear correlation. A value of zero represents no correlation. We will use the Pearson r factor to determine the number of iterations that provide the strongest correlation.

We investigate the application of 3D rrVAE to the simulated graphene data set in Fig. 3. The graphene unit cell used for the simulation is shown in Fig 3 (b). The angle and magnitude of the COM deflection are shown in 3 (c) and 3 (d), respectively. The COM magnitude plot has the expected distribution with minima on the atomic sites and the strongest deflections closest to the atoms. The rotation map in Fig. 3 (e) illustrates the rotations of the CBED patterns about the atomic sites, albeit with reversed polarity. We used 1000 iterations in this case. The latent space observed in Fig. 3 (e) has a strong negative correlation with the COM magnitude map shown in Fig. 3 (d). The second latent space shows a weak correlation and is almost two orders of magnitude smaller in range. A similar trend is observed in Supplementary Fig. S2 where 5 latent dimensions are used. One space has a strong linear correlation with the COM map, but the others have little or no correlation. For completeness the 3D rrVAE analysis of the graphene simulations with temporal and spatial incoherence included are shown in Supplementary Fig. S3. This is essential to get quantitative agreement with the experiment.[30] These smoother results converge in only 300 iterations and both latent spaces show a strong negative correlation. The correlation of one of these spaces degrades rapidly on either side of 300 iterations, while the other space remains relatively stable.

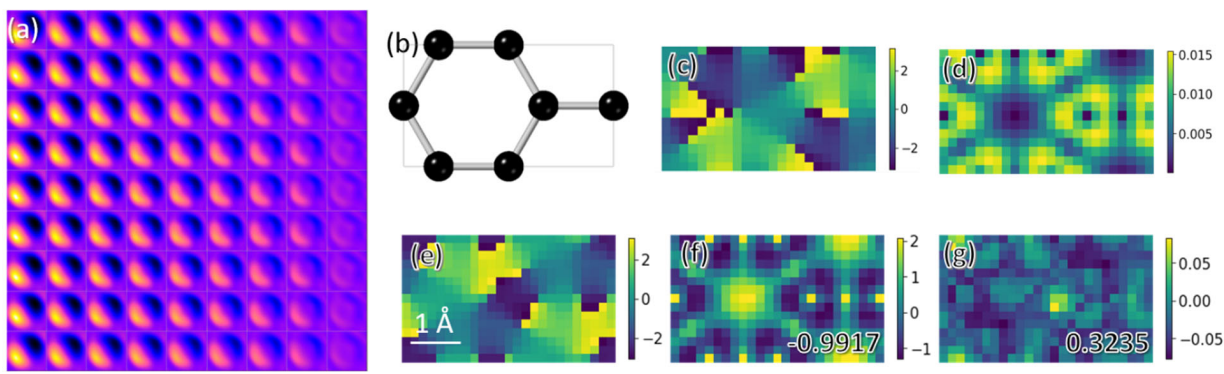

**Figure 3.** (a) Latent space for 3D rrVAE of simulated graphene CBED patterns for microscope operating a 60 kV with a 31 mrad probe forming aperture. (b) Model of the unit cell used for multi-slice simulation. (c) and (d) angle and magnitude of the COM deflection calculated from the simulated CBED patterns, respectively. Scale bar on (e) is in Å$^{-1}$ (e) Rotation map obtained from rrVAE analysis. (f) and (g) show two latent space distributions (with Pearson r factor inset).

Light, 2D materials like graphene represent a special case for 4D-STEM measurements, with very little intensity beyond the bright field center disc of the CBED pattern. For a a more substantial crystalline sample, there is significant intensity beyond this radius. The results of 3D rrVAE on simulated CBED data for ZnS oriented along the [011] zone axis is shown in Fig. 4. The result converges quickly with only 250 iterations. The rotation map shown in Fig. 4 (e) has the opposite polarity to the the angular ditribution of the COM deflection shown in Fig. 4 (c). The latent spaces show a much lower correlation with the COM magnitude than observed for graphene This may be due to the strong asymmetry across the dumbell or perhaps the much stronger scattering in this case.

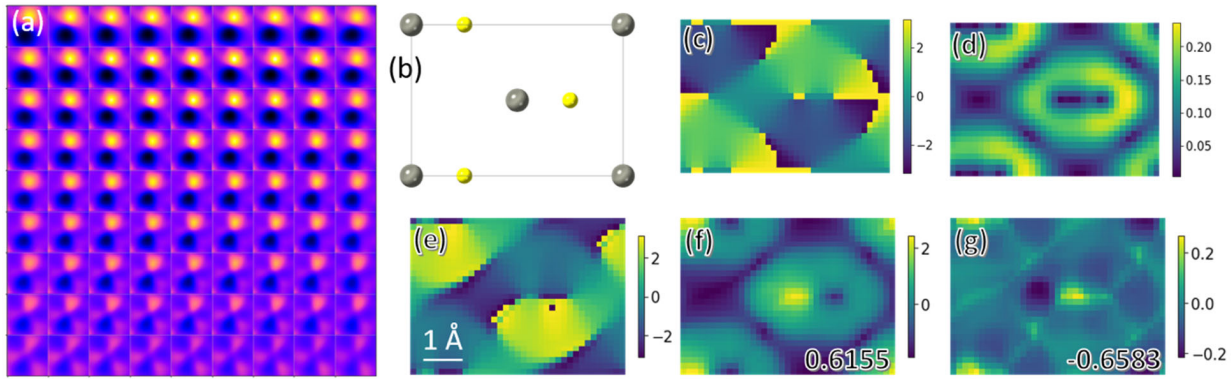

**Figure 4.** (a) Latent space for 3D rrVAE of simulated ZnS [011] zone axis CBED patterns for a microscope operating a 60 kV with a 31 mrad probe forming aperture. A thickness of 76 Å was used in the simulation. Spatial incoherence with a FWHM of 0.75 Å is included. (b) Model of unit cell used for multi-slice simulation. (c) and (d) angle and magnitude of the COM shift calculated from the simulated CBED patterns, respectively. (e) Rotation map obtained from rrVAE analysis. (f) and (g) show two latent space distributions with Pearson r factor inset. 250 iterations were used.

We further extend this approach to an experimental data set. It should be noted that compared to the theoretical data, experimental images have a number of artefacts, including distortion of the image in the probe position (x,y) plane. Since the camera on the Nion UltraSTEM

200 requires relatively long dwell times at each probe position, which accentuates microscope instabilities and drift compared to normal imaging conditions. In addition, the optically coupled camera reveals a bright ring about the edge of the CBED pattern, a distortion that must be addressed before further analysis is possible (two factors contribute to this effect: optical coupling to the scintillator and a condenser-lens dependent effect). The direct application of rrVAE on such data sets often leads to spurious results since the artifacts present in image contrast start to dominate the latent space behaviors.

Several strategies for image rectification based on both the physics of the imaging process and phenomenological exploration were investigated. It was found that subtracting the average CBED intensity over all probe positions, as done previously, removed the spurious distortion around the CBED patterns due to the camera setup. To reduce the size of the rrVAE analysis we binned each CBED image from the as-acquired 256x256 pixels to a more manageable 64x64 pixels. This reduction was a good compromise for the data sets examined here, though each experiment may need to be explored on a case by case basis. This rebinning should perhaps be best applied at the experimental level (on-chip binning usually results in faster possible readout-speeds, reducing the acquisition time). In addition, to reduce noise we applied PCA as implemented in the scikit-learn Python package.[31] An illustrative selection of PCA components from the analysis of experimental graphene CBED patterns are shown in Supplementary Fig. S4.

Using this approach, we applied the rrVAE algorithm to the experimental 4D-STEM data obtained from graphene. This is illustrated in Fig. 5 with the simultaneously acquired annual dark field (ADF)-STEM image shown in Fig. 5 (b) and the COM deflection angle and magnitude, calculated from the processed data, shown in Fig. 5 (c) and 5 (d), respectively. The rotation plot produced by rrVAE is in phase with that derived from the experimental CBED patterns. The latent space shown in Fig. 5 (f) shows a low correlation. The latent space in Fig. 5 (g) has a stronger correlation, but is still quite weak, which is most likely due to the noisy nature of the data. Increasing the number of latent spaces to 5, as shown in Supplementary Fig. S5, does not provide more clarity, although the overall correlation is similar.

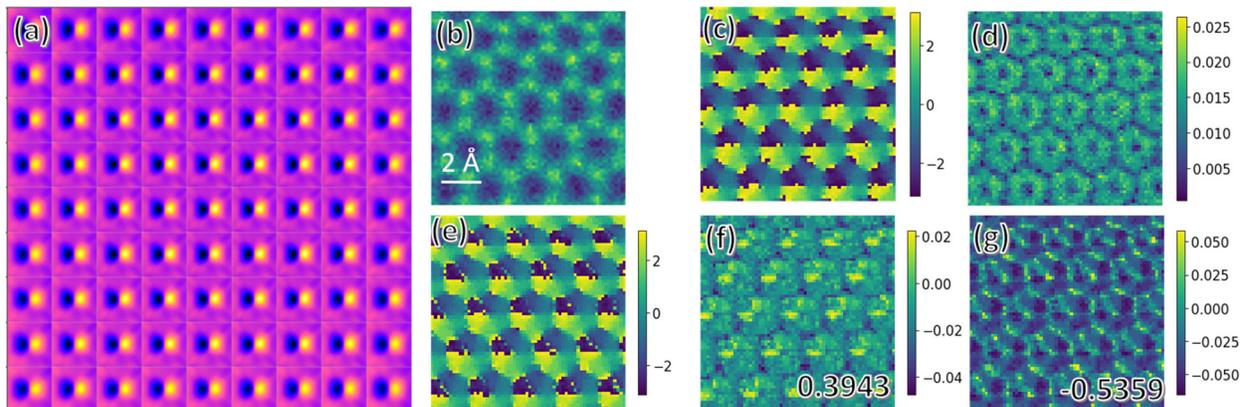

**Figure 5** (a) Latent space for 3D rrVAE for experimental graphene CBED patterns with the microscope operating a 60 kV with a 31 mrad probe forming aperture. (b) Simultaneously acquired ADF-STEM image. (c) and (d) angle and magnitude of the COM deflection calculated from the experimental CBED patterns, respectively. (e) Rotation map obtained from rrVAE analysis. (f) and (g) show two latent space distributions with Pearson r factor inset. 75 iterations were used.

Figure 6 illustrates the effects of defects in graphene over two different length scales. The top row of images show the ADF-STEM, COM, rotation, and latent spaces for graphene with a 3-fold Si impurity over a 1 nm X 1 nm field of view. The Si dopant is obvious in the ADF-STEM image but it is not strong in the COM map. The second latent space is similar to that observed in the pure graphene case in Fig. 5 (f). If 5 latent spaces are used, the degree of correlation is very much reduced, as shown in Supplemental Fig. S6. This is most likely due to the noise level. Interestingly, the position of the Si impurity is highlighted in Fig S6 (c), suggesting a more careful analysis of the latent spaces may yield more than a COM analog.

The lower row of images of Fig. 6 show a vacancy in graphene over a 2 nm X 2 nm field of view. The vacancy is clear in both the ADF-STEM image and COM map. The second latent space has a reasonable correlation with the COM map, but little can be seen in the first latent space. The expansion to 5 latent spaces, as shown in Supplementary Fig. S7, educes the maximum correlation with the defect, which is clearly seen in only one space. In general, the presence of noise is better handled with fewer latent spaces.

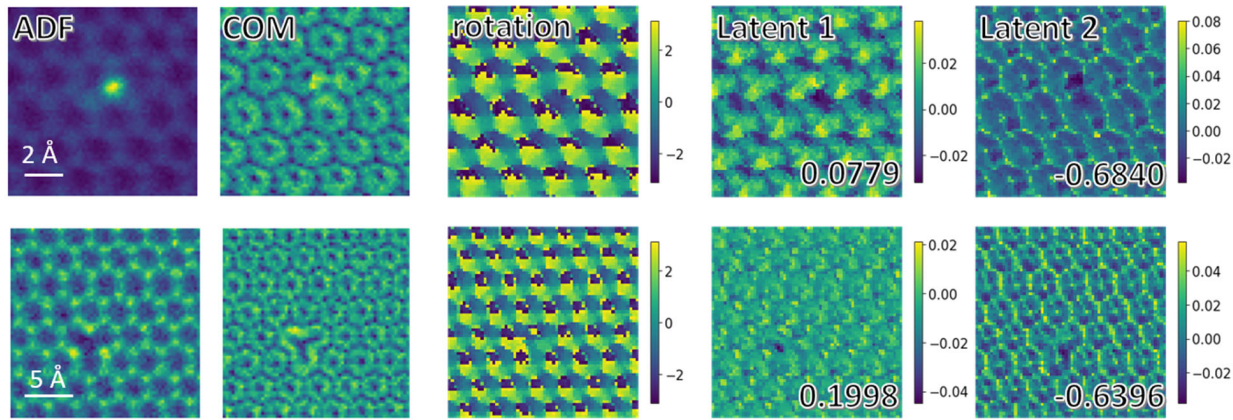

**Figure 6** Top row: Graphene with a 3-fold Si impurity and 1 nm field of view after 90 iterations. Bottom row: Graphene with a defect and 2 nm field of view after 55 iterations.

To summarize, an approach for the analysis of local symmetry breaking via ML analysis of 4D-STEM images has been developed. The rotationally invariant variational autoencoder (rrVAE) approach enables parsimonious representation of the 4D-STEM data in terms of a small number of latent variables including the rotation angle. This approach allows the visualization of the structure of the 4D-STEM data sets in terms of a small number of compact maps, thus directly visualizing symmetry breaking phenomena on the atomic level.

This approach is able to highlight both a single dopant atom and a single vacancy in monolayer graphene. Interestingly, it achieves this result not by examining the high-angle scattered intensity, but through probing the symmetry in the local scattering distribution. This distinction is important because several factors contribute to the ADF-STEM image intensity, making it difficult to distinguish things such as sample thickness changes or surface roughness from intrinsic effects. In the future this method should provide a route to probe defects in cases where there is a small (or no) atomic number difference and to identify visually distinct, but symmetry related, anomalies.

The proposed approach is expected to be universal for the analysis of hyperspectral imaging data sets containing multiple *a priori* unknown rotational variants. As such, it can be directly applied for a broad range of diffraction methods exploring the 2D diffraction spaces of system, including X-ray ptychography, EBSD, and more complex methods. Beyond exploratory image analysis, this approach provides a universal framework for probing symmetry breaking phenomena in complex atomic and mesoscopic systems.

**Acknowledgements:**

This effort (ML and STEM) is based upon work supported by the U.S. Department of Energy (DOE), Office of Science, Basic Energy Sciences (BES), Materials Sciences and Engineering Division (M.P.O., A.R.L., S.V.K., O.D.) and was performed and partially supported (M.Z.) at the Oak Ridge National Laboratory's Center for Nanophase Materials Sciences (CNMS), a U.S. Department of Energy, Office of Science User Facility. This research used resources of the Compute and Data Environment for Science (CADES) at the Oak Ridge National Laboratory, which is supported by the Office of Science of the U.S. Department of Energy under Contract No. DE-AC05-00OR22725.

**Materials and methods**

**Materials:** Atmospheric pressure chemical vapor deposition (AP-CVD)[32] was used to grow graphene on Cu foil. Poly(methyl methacrylate) (PMMA) was spin coated on top of the graphene to protect the surface and form a mechanical stabilizer to facilitate the wet transfer to a TEM grid. The Cu foil was etched away in a bath of ammonium persulfate and deionized (DI) water and the graphene/PMMA stack was rinsed in DI water to remove residues. The graphene/PMMA stack was caught on a TEM grid and baked on a hot plate at 150 °C for 15 min. to promote adhesion of the graphene to the grid. After cooling, the grid was immersed in acetone to dissolve the PMMA and then dipped in isopropyl alcohol to remove the acetone and then dried in air. To remove residual hydrocarbon contaminants the sample was baked in an Ar-$O_2$ atmosphere (10% $O_2$) for 1.5 h at 500 °C.[33] Prior to loading the sample into the STEM, the sample and holder cartridge were baked in vacuum at 160 °C for 8 hours.

**4D STEM measurements:** A Nion UltraSTEM 100 operated at 60 kV was used to acquire the experimental 4D STEM datasets. The CBED images were recorded with an optically coupled Hamamatsu Orca CMOS camera with a 2k by 2k pixel array. The camera was binned to 256 by 256 to increase read out speed. A nominal beam current of 60 pA and nominal convergence angle of 31 mrad was used.

**4D STEM simulation:** All CBED patterns were calculated using a modified version of the μSTEM package[34]. Graphene CBED simulations were carried out using the quantum excitation of phonons algorithm. For the simulations containing incoherence, temporal incoherence was added using weighted sum of defocus values over ±100 Å assuming a Gaussian energy distribution with a full width half maximum (FWHM) of 0.35 eV. Spatial incoherence was added using a weighted sum over CBED patterns and a Gaussian source size with a FWHM of 1.3 Å. Simulations for ZnS were done using the absorptive model and included a source size broadening with a FWHM of 0.75 Å. For the ZnS simulations the probe was focused into the midpoint of the crystal.

# Probing atomic-scale symmetry breaking by rotationally invariant machine learning of multidimensional electron scattering

Mark P. Oxley,[1] Maxim Ziatdinov,[1,2,1] Ondrej Dyck,[1] Andrew R. Lupini,[1] Rama Vasudevan,[1] and Sergei V. Kalinin[1,2]

[1] Center for Nanophase Materials Sciences, Oak Ridge National Laboratory, Oak Ridge, TN 37831, USA

[2] Computational Sciences and Engineering Division, Oak Ridge National Laboratory, Oak Ridge, TN 37831, USA

**Supplementary Figures**

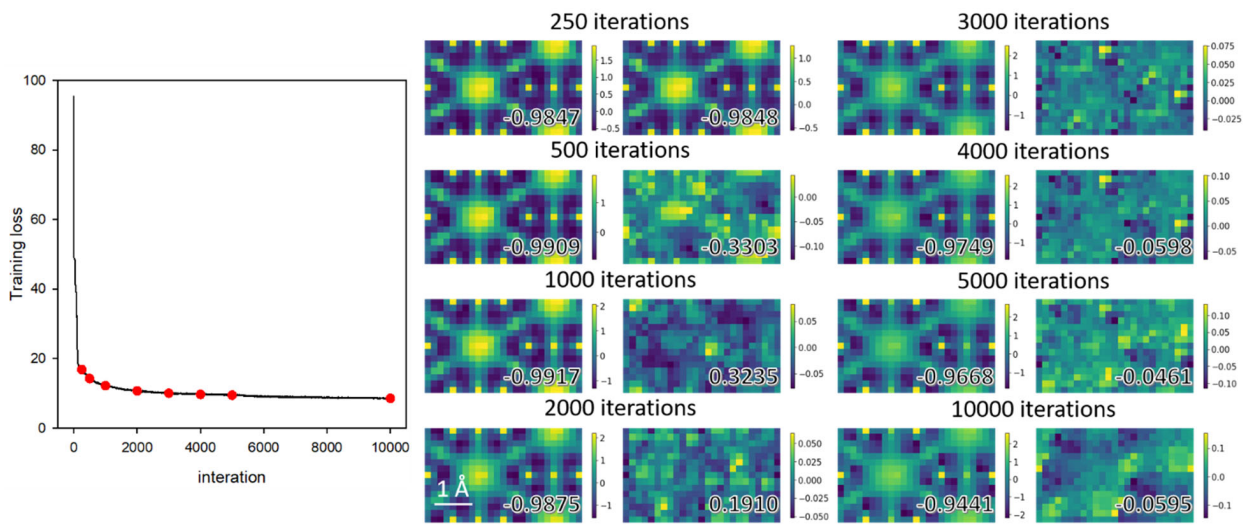

**Figure S1** Convergence of rrVAE for simulated graphene for two latent dimensions. The line plot tracks the training loss as a function of iterations and the red dots represent the points at which the latent spaces are plotted to the right. The inset numbers on each panel show the Pearson correlation coefficient between the latent dimension and the center of mass deflection shown in Fig. 2 of the main text.

[1] ziatdinovm@ornl.gov
[2] sergei2@ornl.gov

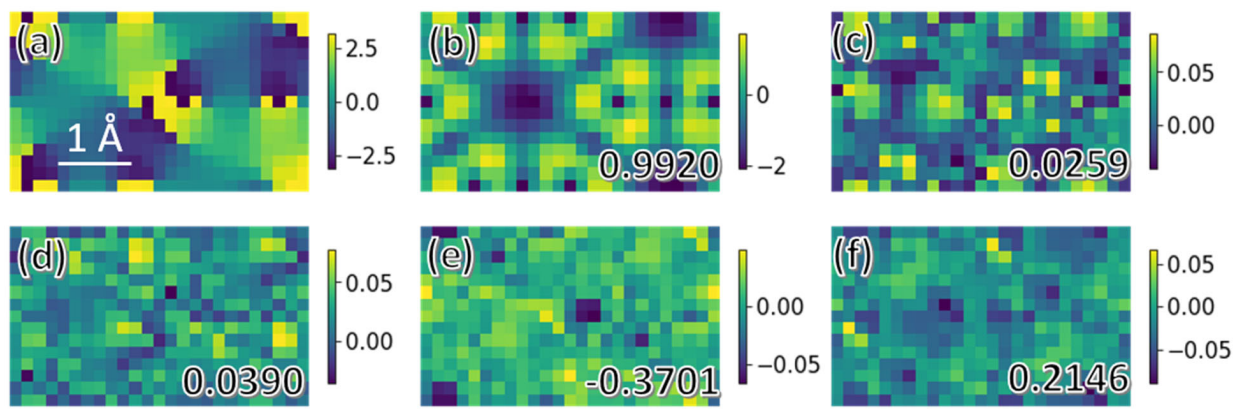

**Figure S2.** Latent space for 6D rrVAE of simulated graphene CBED patterns for a perfect microscope operating a 60 kV with a 31 mrad probe forming aperture. (a) The rotation map obtained from the rrVAE analysis and (b)-(f) the five latent space distributions (with their Pearson r factor inset). 1000 iterations were used for this result.

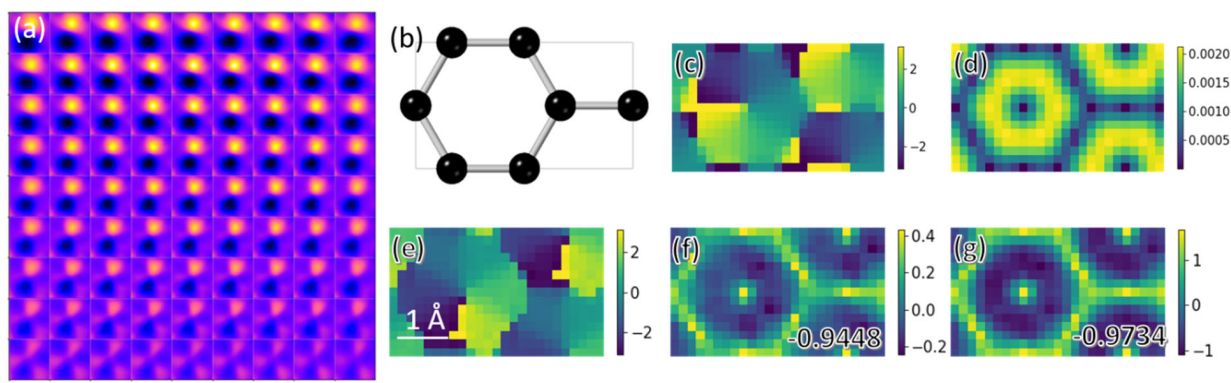

**Figure S3.** (a) Latent space for 2D rrVAE of simulated graphene CBED patterns for a microscope operating a 60 kV with a 31 mrad probe forming aperture with incoherence included in the simulations. (b) The model of the unitcell used for the multislice simulation. (c) and (d) The resulting COM deflection angle and magnitude calculated from the simulated CBED patterns. (e) The rotation map obtained from the rrVAE analysis. (f) and (g) show the two latent space distributions (with their Pearson r factor inset). 300 iterations were used to obtain this result.

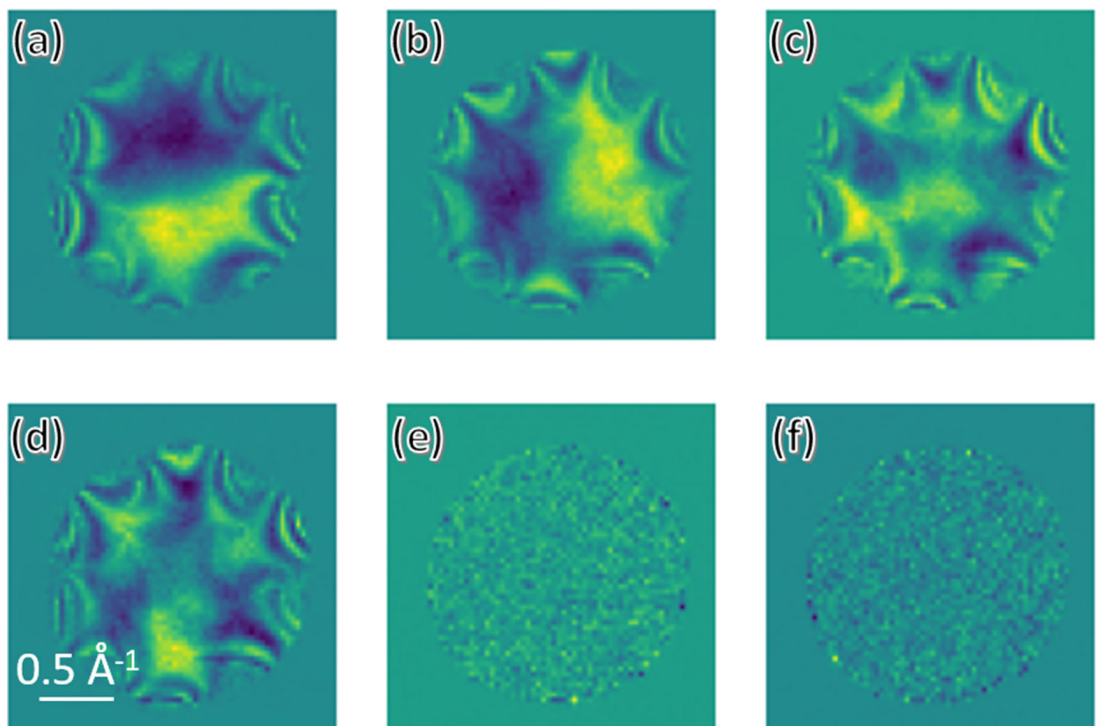

**Figure S4.** A selection of typical PCA components obtained analyzing the experimental graphene CBED patterns. (a)-(d) illustrate the type of components used in the reconstruction. (e) and (f) are typical of the components that are discarded.

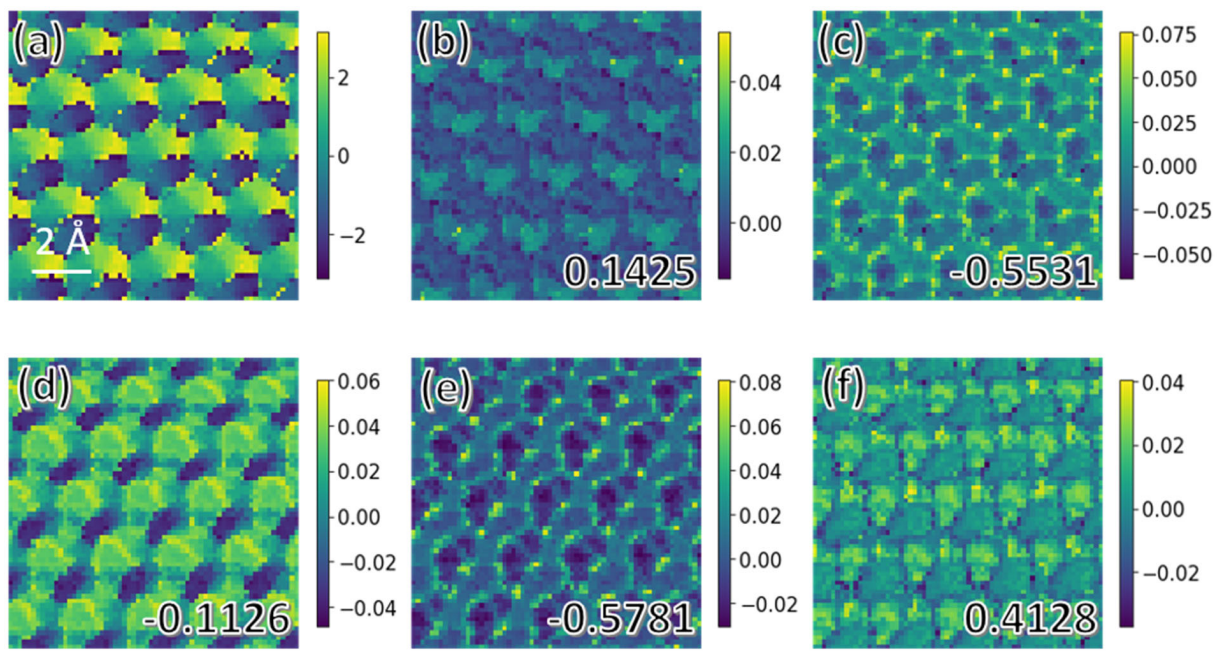

**Figure S5.** Latent space for 6D rrVAE of experimental graphene CBED patterns for a microscope operating a 60 kV with a 31 mrad probe forming aperture. (a) The rotation map obtained from the rrVAE analysis and (b)-(f) the five latent space distributions with Pearson r value inset. 130 iterations were used.

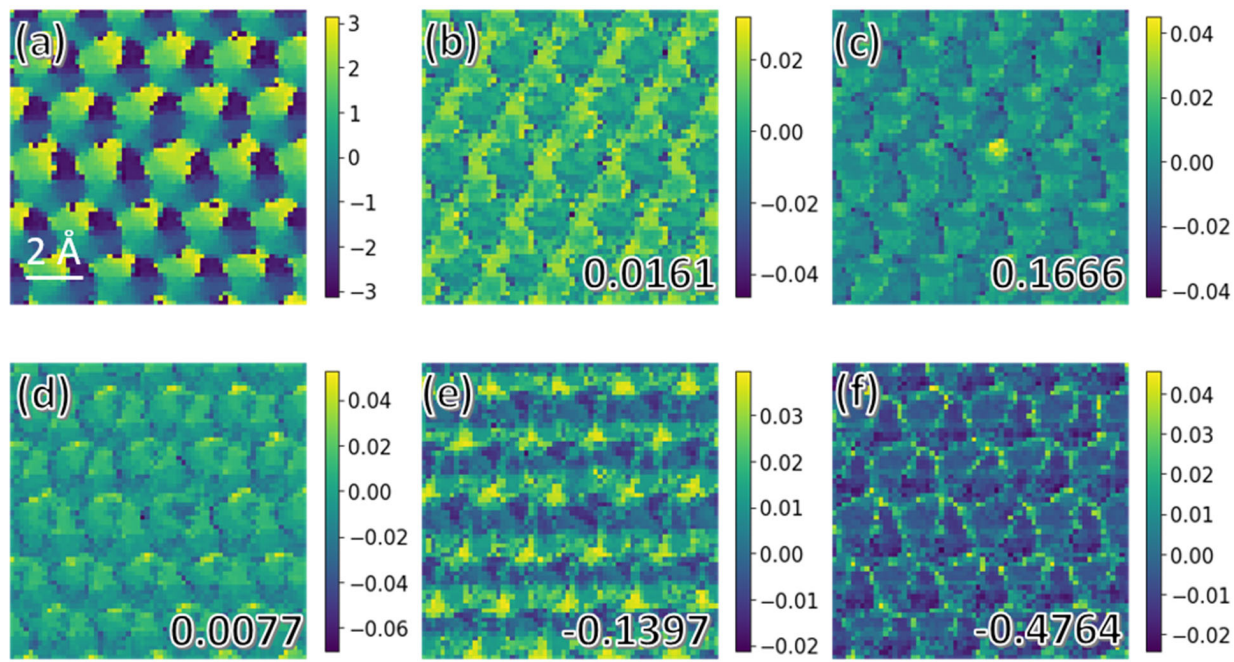

**Figure S6.** Latent space for 6D rrVAE of experimental CBED patterns for graphene with a 3-fold Si impurity for a microscope operating a 60 kV with a 31 mrad probe forming aperture. (a) The rotation map obtained from the rrVAE analysis and (b)-(f) the five latent space distributions with Pearson r value inset. 120 iterations were used

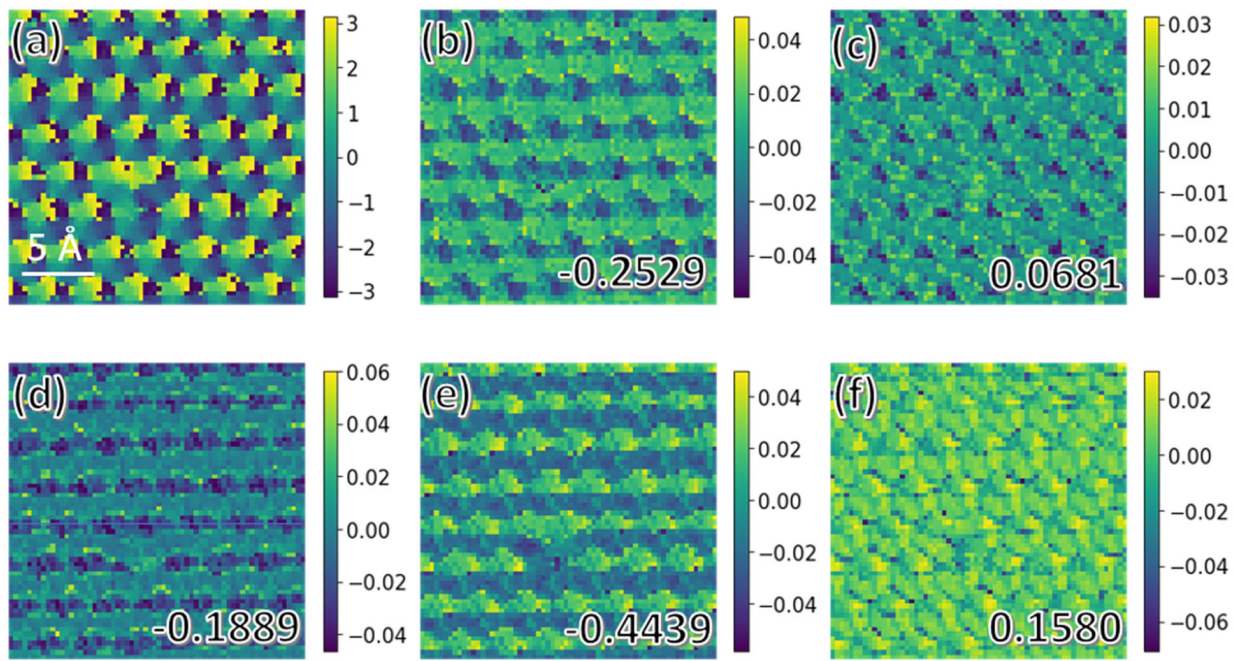

**Figure S7.** Latent space for 6D rrVAE of experimental CBED patterns for graphene with a vacancy for a microscope operating a 60 kV with a 31 mrad probe forming aperture. (a) The rotation map obtained from the rrVAE analysis and (b)-(f) the five latent space distributions with Pearson r value inset. 130 iterations were used.